\documentclass[oneside,12pt]{article}

\usepackage{color}
\usepackage{float}
\usepackage{amsmath,amssymb}
\usepackage{mathrsfs}
\usepackage{enumerate}
\usepackage{breqn}
\usepackage{bm}
\usepackage{harpoon}
\usepackage{mathtools}
   \mathtoolsset{showonlyrefs}
\usepackage{graphicx}
\usepackage{epsfig,psfrag,subfig}
\usepackage{hyperref}
\hypersetup{bookmarks=false, colorlinks=true, linkcolor=black, urlcolor=black, pagebackref=true, pdfstartview={FitH}, bookmarksopenlevel=2 }

\usepackage{epstopdf}
\usepackage{import}

\DeclareMathAlphabet{\mathbf}{OT1}{cmr}{bx}{it}

\definecolor{red}{rgb}{0.9,0,0}
\definecolor{blue}{rgb}{0.019,0.506,0.62}
\definecolor{green}{rgb}{0.0,0.5,0.2}
\definecolor{darkblue}{rgb}{0.2,0.2,0.5}
\definecolor{orange}{rgb}{0.96,0.45,0.184}

\newcommand{\bydef}{\,\raise.050ex\hbox{\rm:}\kern-.025em\hbox{\rm=}\,}
\newcommand{\defby}{=\raise.075ex\hbox{\kern-.325em\hbox{\rm:}}\,}

\def\qed{\relax\ifmmode\hskip2em \Box\else\unskip\nobreak\hskip1em $\Box$\fi}
\newcommand {\Cc}  {\mathcal{C}}
\newcommand {\Dc}  {\mathcal{D}}
\newcommand {\Ec}  {\mathcal{E}}

\newcommand {\Uc}  {\mathcal{U}}

\newcommand {\ab} {\mathbf{a}}
\newcommand {\bb} {\mathbf{b}}

\newcommand {\db} {\mathbf{d}}
\newcommand {\eb} {\mathbf{e}}

\newcommand {\pb} {\mathbf{p}}

\newcommand {\xb} {\mathbf{x}}

\newcommand {\ub} {\mathbf{u}}

\newcommand {\Eb} {\mathbf{E}}

\DeclareMathOperator{\Thz}{\overset{(z)}{\Theta}}
\DeclareMathOperator{\Thc}{\overset{(c)}{\Theta}}

\newtheorem{thm}{Theorem}

\newtheorem{proof}[thm]{Proof}

\begin{document}

\title{\vspace{-1cm} \textbf{A new material property of graphene:\\ the bending Poisson coefficient}}

\author{
Cesare Davini$^1$  \!\!\!\!\! \and Antonino Favata$^2$  \!\!\!\!\! \and Roberto Paroni$^3$ 
}

\date{\today}

\maketitle

\vspace{-1cm}
\begin{center}
	{\small
		$^1$ Via Parenzo 17, 33100 Udine\\
		\href{mailto:cesare.davini@uniud.it}{cesare.davini@uniud.it}\\[8pt]
		$^2$ Department of Structural and Geotechnical Engineering\\
		Sapienza University of Rome, Rome, Italy\\
		\href{mailto:antonino.favata@uniroma1.it}{antonino.favata@uniroma1.it}\\[8pt]

		$^3$ DADU\\
		University of Sassari, Alghero (SS), Italy\\
		\href{mailto:paroni@uniss.it}{paroni@uniss.it}
	}
\end{center}

\pagestyle{myheadings}
\markboth{C.~Davini, A.~Favata, R.~Paroni }
{A new material property of graphene: the bending Poisson coefficient}

\vspace{-0.5cm}
\section*{Abstract}
The in-plane infinitesimal deformations of graphene are well understood: they can be computed by solving the equilibrium problem for a sheet of  isotropic elastic material with suitable stretching stiffness and Poisson coefficient $\nu^{(m)}$.
Here, we pose the following question:
does the  Poisson coefficient $\nu^{(m)}$ affect the response to bending of graphene? Despite  
what happens if one adopts classical structural models, it does not. In this letter we show that a new material property, conceptually and quantitatively different from $\nu^{(m)}$, has to be introduced. We term this new parameter \textit{bending Poisson coefficient}; we propose for it a physical interpretation in terms of the atomic interactions and produce a quantitative evaluation.

\vspace{0.5cm}
\noindent \textbf{PACS}: 61.48.Gh (Structure of graphene), 62.20.dj (Poisson's ratio), 62.20.dq (Other elastic constants)

\tableofcontents

\section{Introduction}

The in-plane energy of an isotropic elastic sheet, occupying a two-dimensional region  $\Omega$,  is
\begin{equation}\label{sheet}
\Uc^{(m)}=\Ec\int_\Omega\nu^{(m)} (\mbox{tr}\, \Eb)^2+(1-\nu^{(m)})|\Eb|^2,
\end{equation}
where, for $\ub$  the in-plane infinitesimal displacement,
$\Eb=1/2 (\nabla\ub+\nabla\ub^T)$ is the strain, $\Ec$ is the \textit{stretching stiffness}, and
$\nu^{(m)}$ is the Poisson coefficient.
Instead, the bending energy of the sheet is
\begin{equation}\label{platen}
\Uc^{(b)}=\frac{1}{2}\,\Dc\int_\Omega(\Delta w)^2-2(1-\nu^{(b)})\det\nabla^2 w,
\end{equation}
where $w$ is the infinitesimal transversal displacement;  $\Dc$ and $\nu^{(b)}$ are two parameters,
the first of which is   the \textit{bending stiffness}.
Energies \eqref{sheet} and \eqref{platen} are classically deduced from a three-dimensional model
from which it results that $\nu^{(b)}=\nu^{(m)}$.

In \textit{Davini}\cite{Davini_2014}, see also references therein,
it has been shown that the energy associated to in-plane infinitesimal deformations of graphene
has the same form of \eqref{sheet} with $\Ec$ and $\nu^{(m)}$ given in terms of elastic lattice constants.

Here, we are mainly concerned with the bending behavior of graphene.
At macroscopic level, graphene can be considered as an elastic sheet able to sustain bending, a feature that  has attracted a growing interest in the past few years  because of the possible technological applications \cite{Ferrari2014}.
In this letter we show that the bending energy associated to infinitesimal displacements of graphene can be recast as in \eqref{platen}, but 
the identity  $\nu^{(b)}=\nu^{(m)}$, which holds for classical structural models, does not hold. Indeed, we show that for a graphene sheet
$$\nu^{(b)}\ne\nu^{(m)}.$$
We call \textit{bending Poisson coefficient} the new independent parameter $\nu^{(b)}$ .

\begin{figure}[h]
	\centering
	\vspace{-1cm}
	\includegraphics[scale=0.9]{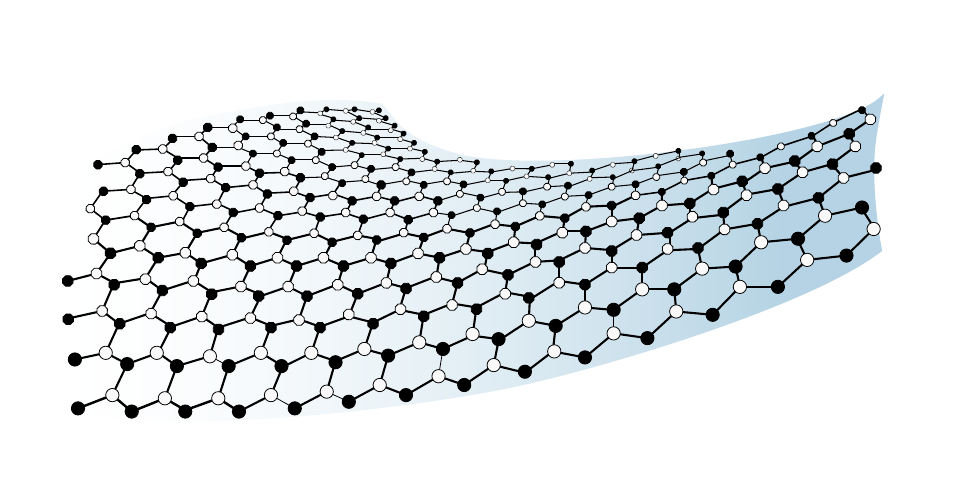}
	\vspace{-1cm}
	\caption{A sheet of graphene in the shape of a double curvature, anticlastic surface.}
	\label{fig:bent_graphene}
\end{figure}

A major conclusion is therefore that the material parameters $\Dc$ and $\nu^{(m)}$, which  are well known in the literature on graphene, are not enough to fully describe the bending behavior. Indeed, when dealing with the small-deformations, the research has so far mainly focused on the in-plane behavior or on cylindrical bending of graphene; in this latter case, as is clear from \eqref{platen}, the bending Poisson coefficient has no role because $\det\nabla^2 w=0$. 

When double curvature deformations are considered (see Fig. \ref{fig:bent_graphene}), it is usually supposed that $\nu^{(b)}=\nu^{(m)}$. In  \textit{Zhang et al.}\cite{Zhang_2011}, the validity of the energy \eqref{sheet}-\eqref{platen} is questioned, basically because of the lack of a specific counterpart at the microscopic level; it is there adopted a  $\pi$-orbital axis vector model and tight binding computations based on  density functional are performed, and some limits on the validity of the standard continuum theory have been described. Nevertheless, in the simulations carried out just cylindrical bending has been considered, a circumstance that cannot highlight the need of considering a bending Poisson coefficient as already noticed. 

The bending stiffness $\Dc$ has been lengthy investigated. An account of the literature on the subject would be necessarily incomplete, and the reader is addressed to the very recent review by \textit{Akinwande et al.}\cite{Akiwande_2017}. In particular, we refer to evaluations based on molecular dynamics computations\cite{Huang2006,Lu_2009,Favata_2016b}. Other techniques, such as density functional theory, tight binding, quantum mechanics-based methods, can be found in the literature \cite{Lindahl_2012,Tapaszto_2012,Kim_2012,Shi_2012,Hartmann_2013,Hajgato_2012,Wei_2013,Pacheco_2014}.  Direct measurement of $\Dc$ has been challenging for monolayer graphene as well as for other 2D materials. The value often quoted for the bending modulus of monolayer graphene is $\simeq 1.2$ eV, estimated from the phonon spectrum of bulk graphite \cite{Nicklow_1972}. \textit{Lindahl et al.}\cite{Lindahl_2012} determined the bending stiffness of double-layer graphene based on measurements of the critical voltage for snap-through of pre-buckled graphene membranes; the same method was applied to monolayer graphene membranes, yielding a rough estimate  with higher uncertainties due to rather limited data point. The membranal Poisson coefficient $\nu^{(m)}$, on the other hand, has been determined by means of density functional theory\cite{Kudin_2001,Wei_2009} and by molecular dynamics simulations\cite{Brenner_1990,Brenner_2002,Stuart2000,Lindsay_2010}.

For the new material constant $\nu^{(b)}$,  we furnish a nano-scale interpretation and propose an analytical formula induced from the 2nd-generation Brenner potential\cite{Brenner_2002},  which is one of the most used  in molecular dynamics simulations of graphene. This parameter rules the attitude of graphene to form surfaces with double curvature, and is fundamental in designing  whatever flexible graphene-based device.

\section{A MD-induced description of graphene bending}\label{model}
Molecular Dynamics (MD) simulations are often employed to study the mechanical behavior of graphene and the 2nd-generation REBO (Reactive Empirical Bond Order) potential is one of the most used empirical potentials. As is known, this potential was originally developed for hydrocarbons by {\it Brenner et al.}\cite{Brenner_2002}, and is able to accommodate  up to third-nearest-neighbor interactions.

\begin{figure}[h]
	\centering
	\vspace{-5mm}
	\includegraphics[scale=0.6]{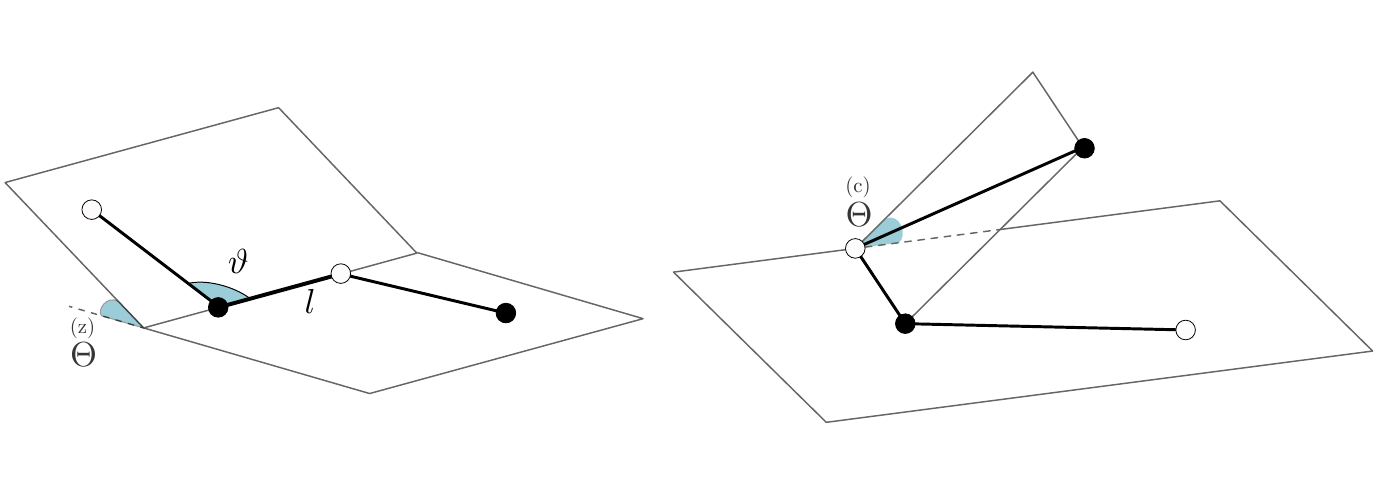}
	\vspace{-5mm}
	\caption{Kinematic variables: distance $l$, angle $\vartheta$, Z-dihedral angle $\Thz$ and C-dihedral angle $\Thc$.}
	\label{fig:bonds}
\end{figure}

In  an easy-to-visualize mechanical picture of graphene,  the kinematic variables associated with the interatomic bonds involve first,  second and third nearest neighbors of any given atom. In particular, the kinematical variables to be  considered are \textit{bond lengths} $l$, \textit{wedge angles} $\vartheta$ and \textit{dihedral angles}. From \textit{Brenner et al.}\cite{Brenner_2002}, dihedral angles are of two kinds that we here term C and Z and denote by $\Thc $ and $\Thz$ (see Fig. \ref{fig:bonds}). The first two contributions are related to the strong covalent $\sigma$-bonds between atoms in one and the same given plane, while the role of dihedral angles is to account for the local coordination of, and the bond angles relative to, two atoms, and are related to the $\pi$-bonds perpendicular to the plane of $\sigma$-bonds.

We consider a harmonic approximation of the stored energy and  assume that it is given by the sum of the following terms:
\begin{equation}\label{eq:ENER 1bis}
\begin{array}{l}
\displaystyle \mathcal{U}_\ell^l = \frac{1}{2} \, \sum_{\mathcal {E}} k^l \, (\delta l)^2 ,\\
\displaystyle\mathcal{U}_\ell^\vartheta = \tau_{0} \sum_{\mathcal {W}}\delta \vartheta +  \frac{1}{2} \, \sum_{\mathcal {W}} k^\vartheta \, (\delta\vartheta)^2,\\
\displaystyle\mathcal{U}_\ell^\Theta = \frac{1}{2} \, \sum_{\mathcal {Z}} k^{\Theta}\, (\delta\Thz)^2+\frac{1}{2} \, \sum_{\mathcal {C}} k^{\Theta}\, (\delta\Thc)^2.
\end{array}
\end{equation}
$\mathcal{U}_\ell^l$, $\mathcal{U}_\ell^\vartheta $ and $\mathcal{U}_\ell^\Theta$ are the energies of the edge bonds, the wedge bonds and the dihedral bonds, respectively; $\tau_0$ is the wedge self-stress, responsible of the cohesive energy. The role of the wedge self-stress is crucial   and its presence is suggested by the shape of the 2nd generation Brenner potential (see \textit{Favata et al.}\cite{Favata_2016} for details). The sums in \eqref{eq:ENER 1bis} extend to
all edges, $\mathcal{E}$, all wedges, $\mathcal{W}$, all Z-dihedra, $\mathcal{Z}$, and all C-dihedra, $\Cc$. The bond constants $k^l$, $k^\vartheta$, and $k^\Theta$  are deduced by making use of the 2nd-generation Brenner potential.

\begin {figure} [!hbtp]
\begin{center}
	\includegraphics [scale=1] {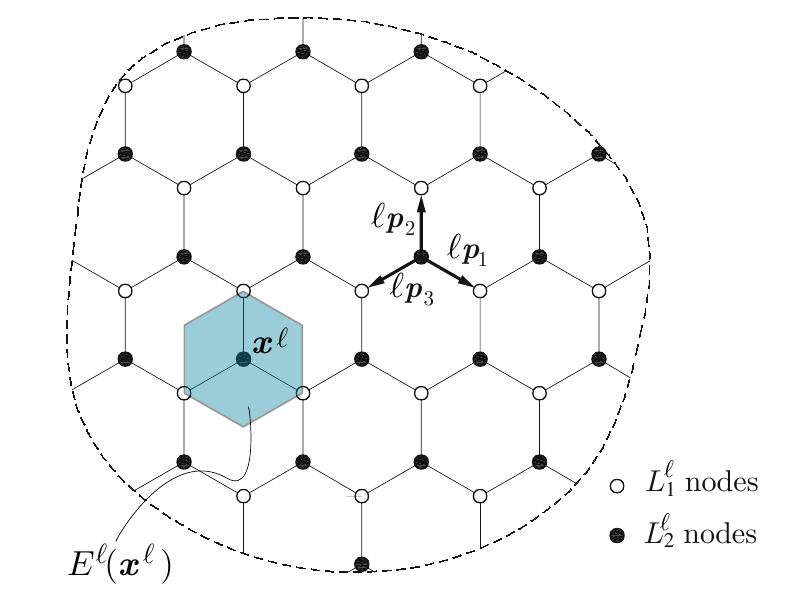}
	\vspace{-5mm}
\end{center}
\caption{The hexagonal lattice.}\label{fg:LATTICE}
\end {figure}

Here, $\ell$ is the natural length of the C--C bond in the reference configuration; $L^\ell_1$ and $L^\ell_2$ denote the two Bravais lattices realizing the graphene layer (see Fig. \ref{fg:LATTICE}, where the nodes of $L^\ell_1$ are represented by blank circles and those of $L^\ell_2$ by black spots); and the attention is restricted to a bounded piece of graphene, that is, to the set of points $\xb^\ell\in L^\ell_1\cup L^\ell_2$ contained in a bounded open set  $\Omega\subset \mathbb{R}^2$, cf. Fig. \ref{fg:LATTICE}. 

In order to express the energies \eqref{eq:ENER 1bis} in terms of the nodal displacements, it is expedient to introduce the  unit vectors  $\pb_i$ ($i=1,2,3$) that, from each node of $L_2^\ell$, point at the nearest neighbor lattice points. 
We also introduce the vectors 
	$$
	\db_1:=\pb_1-\pb_3,\quad\db_2:=\pb_2-\pb_3,\quad\mbox{and}\quad\db_3:=\db_1-\db_2,
	$$ 
	where $\{\ell\db_1,\ell\db_2\}$ is a couple of lattice vectors that generates the lattice $L_1^\ell$.
	We  approximate the strain measures associated to a change of configuration, described by a displacement field $\ub$, by means of a Taylor expansion truncated at the lowest order that makes the energy quadratic in $\ub$. 

We notice that, due to the presence of the wedge self-stress $\tau_0$, also the second order variation of the wedge angles counts. Indeed, if $\delta \vartheta_i^{(1)}(\xb^\ell)$ and $\delta \vartheta_i^{(2)}(\xb^\ell)$ denote the first and the second order variation of the wedge angles at node $\xb^\ell$, respectively, computations that we omit for brevity yield that
$\sum_{i=1}^3 \delta \vartheta_i^{(1)}(\xb^\ell)=0$. Thus, when truncated at the quadratic term, the wedge energy becomes
$$
\displaystyle\mathcal{U}_\ell^\vartheta =\tau_{0} \sum_{\xb^\ell}\sum_{i=1}^3\delta \vartheta_i^{(2)}(\xb^\ell)+\frac{1}{2}  k^\vartheta \, \sum_{\xb^\ell} \sum_{i=1}^3 (\delta \vartheta_i^{(1)}(\xb^\ell))^2,
$$
where, here and below, the sum on $\xb^\ell$ is over the set of points $(L^\ell_2\cup L^\ell_2)\cap\Omega$ unless differently stated.

Computations also show that $\delta l_i$ and $\delta \vartheta_i^{(1)}$  depend upon the in-plane components of $\ub$, while $\delta \vartheta_i^{(2)}$, $\delta\Thc$ and $\delta\Thz$ depend upon the out-of-plane component of $\ub$. This yields a splitting of the energy into \textit{membrane} and \textit{bending} parts 
$$
\mathcal{U}_\ell=\mathcal{U}_\ell^{(m)}+\mathcal{U}_\ell^{(b)},
$$   
with $\mathcal{U}_\ell^{(m)}$ and $\mathcal{U}_\ell^{(b)}$ defined by
$$
\mathcal{U}_\ell^{(m)}:= \frac{1}{2} k^l \, \sum_{\xb_\ell} \sum_{i=1}^3 (\delta l_i(\xb_\ell))^2 + \frac{1}{2}  k^\vartheta \, \sum_{\xb_\ell} \sum_{i=1}^3 (\delta \vartheta_i^{(1)}(\xb^\ell))^2
$$
and
\begin{align}
\mathcal{U}_\ell^{(b)}:=&\tau_{0} \sum_{\xb_\ell}\sum_{i=1}^3\delta \vartheta_i^{(2)}(\xb^\ell)\nonumber\\
&\hspace{3mm}+\frac{1}{2} \, \sum_{\mathcal {Z}} k^{\Theta}\, (\delta\Thz)^2+\frac{1}{2} \, \sum_{\mathcal {C}} k^{\Theta}\, (\delta\Thc)^2.\label{bendergy}
\end{align}

From a detailed study of kinematics, it turns out that the expressions giving the in-plane strains  $\delta l_i$ and $\delta \vartheta_i^{(1)}$ coincide with those worked out by \textit{Davini}\cite{Davini_2014} in dealing with the plane deformations of a graphene sheet. Therefore, that analysis applies and provides a characterization of the continuum limit, for $\ell\to 0$, of the membrane energy $\mathcal{U}_\ell^{(m)}$. So,  hereafter we concentrate on the bending energy only.

The first term on the right hand side of \eqref{bendergy} 
\begin{equation}\label{selfen1}
\mathcal{U}^{(s)}_\ell:= \tau_0\sum_{\xb^\ell}  \sum_{i = 1}^3\delta \vartheta_i^{(2)} (\xb^\ell),
\end{equation}
is the \textit{cohesive energy}. 

\begin{figure}[h]
	\centering
	\includegraphics[scale=0.8]{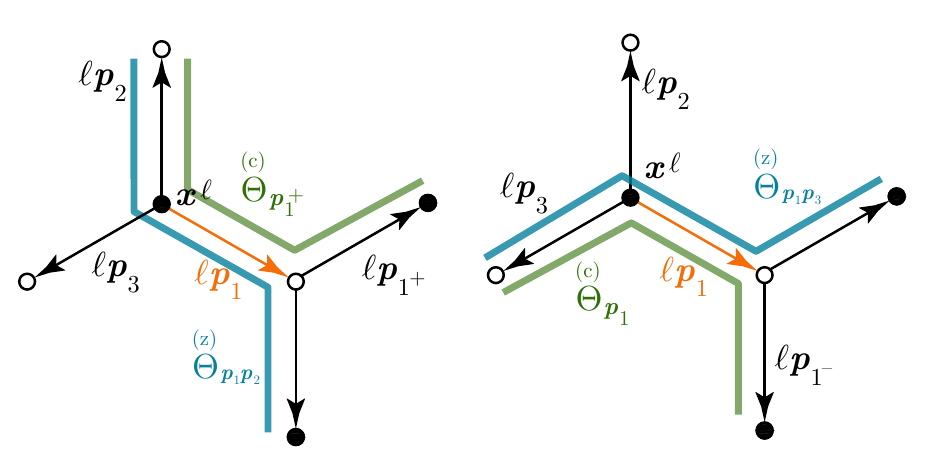}
	\vspace{-5mm}
	\caption{Left: C-dihedral angles $\Thc_{\pb_1^+}$ (green) and Z-dihedral angle $\Thz_{\pb_1\pb_2}$ (blue). Right: C-dihedral angles $\Thc_{\pb_1^-}$ (green) and Z-dihedral angle $\Thz_{\pb_1\pb_3}$ (blue).}
	\label{fig:cell_text}
\end{figure}

Fig. \ref{fig:cell_text} illustrates the meaning of the dihedral angles $\Thz_{\pb_i\pb_{i+2}}$, $\Thz_{\pb_i\pb_{i+1}}$, $\Thc_{\pb_i^+}$, $\Thc_{\pb_i^-}$, defined as the angles between the planes spanned by  two adjacent segments in colored chain; with these, the other two terms in the energy can be written as
\begin{align}\label{enZ}
\mathcal{U}^{\mathcal{Z}}_\ell=\frac{1}{2}\, k^{\Theta} \sum_{\xb^\ell\in L^\ell_2\cap\Omega}
\sum_{i=1}^3 &\Bigg(\delta\Thz_{\pb_i\pb_{i+2}}(\xb^\ell)\Bigg)^2+\\
&+ \Bigg(\delta\Thz_{\pb_i\pb_{i+1}}(\xb^\ell)\Bigg)^2,\nonumber
\end{align}
and 
\begin{equation}\label{enC}
\mathcal{U}^{\mathcal{C}}_\ell=\frac{1}{2} k^{\Theta} \, \sum_{\xb^\ell\in L^\ell_2\cap\Omega}
\sum_{i=1}^3 \Bigg(\delta\Thc_{\pb_i^+}(\xb^\ell)\Bigg)^2+ \Bigg(\delta\Thc_{\pb_i^-}(\xb^\ell)\Bigg)^2,
\end{equation}
where $\mathcal{U}^{\mathcal{Z}}_\ell$  and  $\mathcal{U}^{\mathcal{C}}_\ell$ respectively are the {\it Z-dihedral energy} and the {\it C-dihedral energy}.   Here, $i, i+1, i+2$ take values in $\{1,2,3\}$ and the sum is to be interpreted modulo 3: for instance, if $i=3$, then $i+1=1$ and $i+2=2$.

In \eqref{enZ} and \eqref{enC} we have taken the elastic constants of both types of dihedral energy to be equal, as  implied by the potential adopted. We notice also that the term $\sum_{i=1}^3\delta \vartheta_i^{(2)} (\xb^\ell)$ in the cohesive energy, cf. \eqref{selfen1},  is always non-positive, as is made clear in the Appendix. Hereafter, we assume that  
	$$
	\tau_0<0\quad\mbox{and}\quad k^{\Theta} >0,
	$$
	so that all the components of the bending energy are non-negative.

\section{The continuum bending energy}
The discrete bending energy $\mathcal{U}_\ell^{(b)}$ depends on out-of-plane displacements defined over the set of points  $(L^\ell_1\cup L^\ell_2)\cap\Omega$. We assume that it can be approximated by a continuous energy $\Uc_0^{(b)}$ depending on functions defined over the domain $\Omega$, to be determined by letting the lattice size $\ell$ go to zero so that
$(L^\ell_1\cup L^\ell_2)\cap\Omega$ invades $\Omega$.

Lengthy computations, that we  omit for the sake of  space, yield:
\begin{align}
\delta\Thz_{\pb_i\pb_{i+2}}(\xb^\ell)&=\frac{2\sqrt{3}}{3\ell}[w(\xb^\ell+\ell\pb_i-\ell \pb_{i+2})-w(\xb^\ell)\\
&-w(\xb^\ell+\ell \pb_{i})+w(\xb^\ell+\ell \pb_{i+2})]\nonumber\\
&=\frac{2\sqrt{3}}{3\ell}[w(\xb^\ell+\ell\ab_i)-w(\xb^\ell)\\
&-w(\xb^\ell+\ell \ab_{i}+\ell \pb_{i+2})+w(\xb^\ell+\ell \pb_{i+2})]\nonumber\\
&=\frac{2\sqrt{3}}{3\ell} [- \partial^2_{\ab_i\pb_{i+2}}w(\xb^\ell) \ell^2 |\ab_i| |\pb_{i+2}|+o(\ell^2)]\nonumber\\
&=-2 \ell \partial^2_{\ab_i\pb_{i+2}}w(\xb^\ell) +o(\ell),\label{angZ1}
\end{align}
where $w$ is the out-of-plane  component of displacement  and $\ab_i:=\pb_i-\pb_{i+2}$, $|\ab_i |=\sqrt{3}$; $\partial^2_{\ab_i\pb_{i+2}}$ denotes the second partial derivative in the directions of $\ab_i$ and $\pb_{i+2}$; and $o(\cdot)$ stands for an infinitesimal quantity of order greater than its argument. 
With \eqref{enZ} in mind, we rewrite this identity as
\begin{equation}
\begin{aligned}
\left(\delta\Thz_{\pb_i\pb_{i+2}}(\xb^\ell)\right)^2
=\\
\frac{8\sqrt{3}}9 &\int_{E^\ell(\xb^\ell)}(\partial^2_{\ab_i\pb_{i+2}}w(\xb^\ell))^2\,d\xb+o(\ell^2),\label{angZ2}
\end{aligned}
\end{equation}
where $E^\ell(\xb^\ell)$ is  the hexagon of side $\ell$ centred at $\xb^\ell$,
whose area is $\ell^23\sqrt{3}/2$  (see Fig. \ref{fg:LATTICE}). Similarly, we have:
\begin{equation}
\begin{aligned}
\left(\delta\Thz_{\pb_i\pb_{i+1}}(\xb^\ell)\right)^2
=\\
\frac{8\sqrt{3}}9 &\int_{E^\ell(\xb^\ell)}(\partial^2_{\bb_i\pb_{i+1}}w(\xb^\ell))^2\,d\xb +o(\ell^2),
\end{aligned}
\end{equation}
where $\bb_i:=\pb_i-\pb_{i+1}$.  By writing $\ab_i$ and $\bb_i$ explicitly, using \eqref{enZ}, and observing that the number of items in the sums has order $1/\ell^2$, for $\ell$ going to zero we find that
\begin{align}
\mathcal{U}^{\mathcal{Z}}_0(w)&:=\lim_{\ell\to 0}\mathcal{U}^{\mathcal{Z}}_\ell=\frac{1}{2}\frac{8\sqrt{3}}9 k^{\Theta}\int_\Omega (\partial^2_{\db_1\pb_{3}}w)^2+ (\partial^2_{\db_1\pb_{1}}w)^2\label{limenZ}\\
&\hspace{-5mm}+(\partial^2_{\db_2\pb_{2}}w)^2
+(\partial^2_{\db_2\pb_{3}}w)^2
+(\partial^2_{\db_3\pb_{1}}w)^2+(\partial^2_{\db_3\pb_{2}}w)^2\,d\xb.\nonumber
\end{align}
Hence, by rewriting the directional derivatives in terms of the partial derivatives with respect to an orthogonal system of coordinates $\{x_1, x_2\}$ we deduce that
\begin{align}
\mathcal{U}^{\mathcal{Z}}_0(w)
&=\frac{1}{2}\frac{5\sqrt{3}}3 k^{\Theta}\int_\Omega  (\Delta w)^2- \frac{8}5 \det \nabla^2 w\,d\xb.
\label{enZ0}
\end{align}
Analogous expressions can be derived for $\mathcal{U}^{\mathcal{C}}_0(w)$ and $\mathcal{U}^{(s)}_0(w)$, cf. Appendix. Then, by summing all contributions up, we find that the continuum limit of the bending energy is given by:
\begin{align}\label{entot}
\mathcal{U}^{(b)}_0(w)
&=\frac 12 \Bigg(\frac{7\sqrt{3}}3 k^{\Theta}-\frac{\tau_0}2\Bigg) \int_\Omega( \Delta w)^2\nonumber\\
&\hspace{0.5cm}-2\,\frac{16 k^\Theta}{14 k^\Theta-\sqrt{3}\tau_0}\det \nabla^2 w\,d\xb\nonumber\\
&=\frac 12 \Dc  \int_\Omega ( \Delta w)^2-2(1-\nu^{(b)})\det \nabla^2 w \,d\xb,
\end{align}

where we have set
\begin{equation}\label{nub}
\nu^{(b)}=1-\frac{16 k^\Theta}{14 k^\Theta-\sqrt{3}\tau_0}.
\end{equation}
We call $\nu^{(b)}$ the\textit{ bending Poisson coefficient}.

In sheets of conventional materials, the coefficient $\nu^{(b)}$ appearing in eq. \eqref{entot} is the  membranal Poisson coefficient $\nu^{(m)}$.  For graphene this is no longer so. In fact from the continuum in-plane energy of graphene one finds:
\begin{equation}\label{numm}
\nu^{(m)}=\frac{k^l\ell^2-6k^\vartheta}{k^l\ell^2+18k^\vartheta}\neq\nu^{(b)},
\end{equation}
(see the paper by \textit{Davini}\cite{Davini_2014} for a detailed computation).
Besides the quantitative evaluations, detailed in the next section, it is worth noticing that the two coefficients are conceptually different, as the nano-scale interpretation given by \eqref{nub} and \eqref{numm} clearly reveals. In fact, the membranal Poisson coefficient  depends upon
\begin{enumerate}[(i)]
	\item the reluctance of two arbitrary atoms to change their distance, captured by $k^l$; 
	\item the reluctance of three atoms to change their mutual angle, captured by $k^\vartheta$.
\end{enumerate}	 
The  bending Poisson coefficient, on the contrary, is determined by 
\begin{enumerate}[(i)]
	\item the reluctance of four arbitrary atoms to change the dihedral angles they form, captured by $k^\Theta$,
	\item the amount of the cohesive energy, related to $\tau_0$.
\end{enumerate}

Equation \eqref{entot}  also confirms that the origin of the bending stiffness
\begin{equation}\label{bendstiff}
\Dc=\frac{7\sqrt{3}}3 k^{\Theta}-\frac{\tau_0}2
\end{equation}
is twofold: a part depends on the dihedral contribution, and a part on the cohesive energy (cf. \textit{Favata et al.}\cite{Favata_2016b}). 
Computations following from the quantitative evaluation of $k^{\Theta}$ and $\tau_0$  in the next section show that the two terms contribute to $\mathcal D$ almost equally.

%

\section{Results}
In order to obtain proper quantitative evaluations of the bending Poisson coefficient, let us recall the 2-nd generation Brenner potential\cite{Brenner_2002}.
The  binding energy $V$ of an atomic aggregate is given as a sum over nearest neighbors:
\begin{equation}\label{V}
V=\sum_i\sum_{j<i} V_{ij}\,;
\end{equation}
the interatomic potential $V_{ij}$ is given by 
\begin{equation}\label{Vij}
V_{ij}=V_R(r_{ij})+b_{ij}V_A(r_{ij}),
\end{equation}
where the individual effects of the \emph{repulsion} and \emph{attraction functions} $V_R(r_{ij})$ and $V_A(r_{ij})$, which model pair-wise interactions of  atoms $i$ and $j$ depending on their distance $r_{ij}$, are modulated by the \emph{bond-order function} $b_{ij}$, which depends on the angles $\vartheta_{ijk}$ and the dihedral angles $\Theta_{ijkl}$.
The values of the constant $k^\Theta$  can be then deduced by deriving twice the potential, and computing the result in the ground state, where $r_{ij}=\ell$, $\theta_{ijk}=2/3\pi$, $\Theta_{ijkl}=0$. 
Moreover, from \textit{Favata et al.}\cite{Favata_2016}, we take the value of the selfstress $\tau_0$:
\begin{equation}\label{tau0}
\tau_0=-0.2209\;\textrm{nN nm}=-1.3787\;\textrm{eV}.
\end{equation}
It turns out that the bending Poisson coefficient is:
\begin{equation}
\nu^{(b)}=0.419.
\end{equation}
With the values of the constants $k^l$ and $k^\vartheta$ deduced from the 2nd-generation Brenner potential, we find that the membranal Poisson coefficient \eqref{numm} is
\begin{equation}
\nu^{(m)}=0.397.
\end{equation}

This estimate agrees with the values reported in the literature.  For instance, with results obtained by MD simulations that adopt the 2nd-generation Brenner potential \cite{Arroyo2004,Akiwande_2017}.  For a further check, we can compare the value of the bending stiffness  obtained from \eqref{bendstiff} with others found in the literature. Indeed, from \eqref{bendstiff} we get that $\Dc=1.4022$ eV, which is in good agreement with the results given in references\cite{LuJ1997, Wei_2013,Akiwande_2017}, although they are deduced by completely different approaches.

\section{Conclusions}
In summary, based on the description of  atomic interactions provided by the 2nd-generation Brenner potential, we have found a new material parameter that influences  the bending  behavior of graphene.  This parameter, called bending Poisson coefficient, is conceptually and quantitatively different from the already known membranal Poisson coefficient. We have proposed an analytical evaluation in terms of atomistic quantities, revealing its nano-scale physical sources.

\section*{Acknowledgments}
AF acknowledges the financial support of Sapienza University of Rome (Progetto d'Ateneo 2016 --- ``Multiscale Mechanics of 2D Materials: Modeling and Applications'').

\appendix

	\section{Appendix: Deduction of  $\mathcal{U}^{\mathcal{C}}_0$ and $\mathcal{U}^{(s)}_0$}\label{appendix}
	
	For completeness, here we sketch the deduction of the expressions of $\mathcal{U}^{\mathcal{C}}_0$ and $\mathcal{U}^{(s)}_0$.
	
	By Taylor expansion theorem, we calculate the change of the $\mathcal{C}$-dihedral angle $\delta\Thc_{\pb_i^+}(\xb^\ell)$:
	\begin{align}\label{thC+}
	\delta\Thc_{\pb_i^+}(\xb^\ell)&=\frac{2\sqrt{3}}{3\ell}[2w(\xb^\ell)-w(\xb^\ell+\ell \pb_{i+1})\nonumber\\
	&\hspace{10mm}+w(\xb^\ell+\ell\pb_i-\ell \pb_{i+2})-2w(\xb^\ell+\ell \pb_{i})]\nonumber\\
	&={2\ell}\partial^2_{\pb_i\pb_i^\perp}w(\xb^\ell)+o(\ell),
	\end{align}
	where $\pb_i^\perp:=\eb_3\times\pb_i$. Similarly, we find that
	\begin{equation}\label{thC-}
	\delta\Thc_{\pb_i^-}(\xb^\ell)={2\ell}\partial^2_{\pb_i\pb_i^\perp}w(\xb^\ell)+o(\ell).
	\end{equation}
	
	With \eqref{enC} in mind, we write the identities
	\begin{equation}
	\begin{array}{l}
	\displaystyle\left(\delta\Thc_{\pb_i^+}(\xb^\ell)\right)^2=\frac{8\sqrt{3}}{9}\int_{E^\ell(\xb^\ell)}(\partial^2_{\pb_i\pb_i^\perp}w(\xb^\ell))^2\,d\xb +o(\ell^2)\\
	\displaystyle\left(\delta\Thc_{\pb_i^-}(\xb^\ell)\right)^2=\frac{8\sqrt{3}}{9}\int_{E^\ell(\xb^\ell)}(\partial^2_{\pb_i\pb_i^\perp}w(\xb^\ell))^2\,d\xb +o(\ell^2)
	\end{array}
	\end{equation}
	and deduce that
	$$
	\mathcal{U}^{\mathcal{C}}_\ell=\frac{8\sqrt{3}}{9} k^{\Theta} \sum_{\xb^\ell\in L^\ell_2\cap\Omega}\int_{E^\ell(\xb^\ell)} \sum_{1=1}^3(\partial^2_{\pb_i\pb_i^\perp}w(\xb^\ell))^2\,d\xb+o(1).
	$$
	By passing to the limit we find
	\begin{equation}\label{limenC2}
	\mathcal{U}^{\mathcal{C}}_0(w):=\lim_{\ell\to 0}\mathcal{U}^{\mathcal{C}}_\ell=\frac{8\sqrt{3}}{9} k^{\mathcal{C}}\int_\Omega \sum_{i=1}^3 \Big(\partial^2_{\pb_i\pb_i^\perp}w\Big)^2\,d\xb.
	\end{equation}
	
	We now compute the limit of the  cohesive energy \eqref{selfen1}. 
	
	We focus on $\xb^\ell\in L_2^\ell$, since $\xb^\ell\in L_1^\ell$ produces the same result with the same steps. 
	Computations show that $\delta \vartheta_i^{(2)} (\xb^\ell)$ has the form
	\begin{equation}\label{dth2}
	\sum_{i=1}^3\delta\vartheta_i^{(2)}(\xb^\ell)=-\frac{3\sqrt{3}}{\ell^2}\Bigg(\frac 13 \sum_{i=1}^3w(\xb^\ell+\ell\pb_i)-w(\xb^\ell)\Bigg)^2.
	\end{equation}
	So, in particular, $\delta \vartheta_i^{(2)} (\xb^\ell)$ is non-positive definite.
	
\begin{figure}[h]
		\centering
		\includegraphics[scale=0.9]{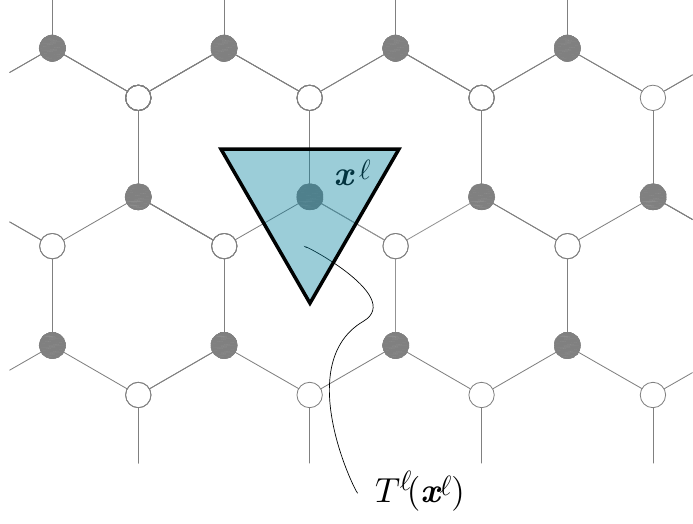}
		\caption{Triangulation  $T^\ell(\xb^\ell)$}
		\label{fig:triangle}
	\end{figure}

	By using Taylor's expansion in \eqref{dth2}, we find
	\begin{align}\label{sumthi}
	\sum_{i=1}^3\delta\vartheta_i^{(2)}(\xb^\ell)&=-\frac{\sqrt{3}}{3} {\ell^2}\Big( \partial^2_{\pb_1 \pb_1}w(\xb^\ell)\nonumber\\
	&\hspace{5mm}+\partial^2_{\pb_2 \pb_2}w(\xb^\ell)+\partial^2_{\pb_1 \pb_2}w(\xb^\ell)\Big)^2+o(\ell^2),
	\end{align}
	that can be written as
	\begin{align}\label{sumthi2}
	\sum_{i=1}^3\delta\vartheta_i^{(2)}(\xb^\ell)&=-\frac{4}{9}\int_{T^\ell(\xb^\ell)}\Big( \partial^2_{\pb_1 \pb_1}w(\xb^\ell)+\partial^2_{\pb_2 \pb_2}w(\xb^\ell)\nonumber\\
	&\hspace{20mm}+\partial^2_{\pb_1 \pb_2}w(\xb^\ell)\Big)^2\,dx+o(\ell^2),
	\end{align}
	where  $T^\ell(\xb^\ell)$ is a triangle centered at  $\xb^\ell\in L_1^\ell\cup L_2^\ell$, with vertices in the center of the hexagonal cells having $\xb^\ell$ in common, see Fig.~\ref{fig:triangle}. Note that the area of $T^\ell(\xb^\ell)$ is $\frac{3\sqrt{3}}4\ell^2$.
	
	With this expression the cohesive energy \eqref{selfen1} takes the form
	\begin{align}
	\mathcal{U}^{(s)}_\ell
	&=-\frac{4}{9}\tau_0\sum_{\xb^\ell\in (L_1^\ell\cup L_2^\ell)\cap\Omega}   
	\int_{T^\ell(\xb^\ell)}\Big( \partial^2_{\pb_1 \pb_1}w(\xb^\ell)\nonumber\\
	&\hspace{10mm}+\partial^2_{\pb_2 \pb_2}w(\xb^\ell)+\partial^2_{\pb_1 \pb_2}w(\xb^\ell)\Big)^2\,dx+o(1). \nonumber 
	\end{align}
	Thence, by passing to the limit, we get
	\begin{align}\label{limentheta2}
	\mathcal{U}^{(s)}_0(w):=&\lim_{\ell\to 0}\mathcal{U}^{(s)}_\ell\nonumber\\
	=&-\frac{4}9\tau_0 \int_\Omega ( \partial^2_{\pb_1 \pb_1}w+\partial^2_{\pb_2 \pb_2}w+\partial^2_{\pb_1 \pb_2}w\Big)^2\, d\xb.
	\end{align}

\end{document}